\documentstyle[twoside,fleqn,espcrc2,epsfig]{article}

\newcommand{\AmS}{{\protect\the\textfont2
  A\kern-.1667em\lower.5ex\hbox{M}\kern-.125emS}}

\newcommand\beq{\begin{equation}}
\newcommand\eeq{\end{equation}}
\newcommand\bea{\begin{eqnarray}}
\newcommand\eea{\end{eqnarray}}
\newcommand\MSbar{$\overline{\mbox{MS}}$}

\newcommand\nnn{\nonumber \\}
\newcommand\nn{\nonumber}
\title{Theoretical Uncertainties in the QCD Evolution of Structure 
 Functions and their Impact on $\alpha_s(M_Z^2) ${\thanks{Talk 
 presented by S. Riemersma}}}
\author{
J. Bl\"umlein$^{\,\rm a}$, S. Riemersma 
 \address{DESY--Zeuthen, Platanenallee 6, D--15735 Zeuthen, Germany},
W.L. van Neerven \address{Instituut Lorentz, Rijksuniversiteit Leiden, 
 PO Box 9506, 2300 RA Leiden, The Netherlands},
and A. Vogt \address{Institut f\"ur Theoretische Physik, Universit\"at 
 W\"urzburg, Am Hubland, D--97074 W\"urzburg, Germany}
}

\begin{document}
\begin{titlepage}

\large
\begin{flushleft}
DESY 96--172 \\[0.1cm] 
INLO-PUB-17/96 \\[0.1cm]
WUE-ITP-96-017 \\[0.1cm]
August 1996
\end{flushleft}
\vspace{0.4cm}
\begin{center}
\LARGE
{\bf Theoretical Uncertainties in the QCD Evolution of }\\
\vspace{0.1cm}
{\bf Structure
 Functions and their Impact on $\alpha_s(M_Z^2)$} \\
\vspace{1.0cm}
\large
J. Bl\"umlein, S. Riemersma \\
\vspace{0.4cm}
\large {\it
DESY--Zeuthen \\
\vspace{0.1cm}
Platanenallee 6, D--15735 Zeuthen, Germany }\\
\vspace{0.8cm}
W.L. van Neerven \\
\vspace{0.4cm}
\large {\it
Instituut Lorentz, Rijksuniversiteit Leiden,}\\
\vspace{0.1cm}
{\it
 PO Box 9506, 2300 RA Leiden, The Netherlands}\\
\vspace{0.8cm}
\large
A. Vogt\\
\vspace{0.4cm}
\large {\it
Institut f\"ur Theoretische Physik, Universit\"at W\"urzburg \\
\vspace{0.1cm}
Am Hubland, D--97074 W\"urzburg, Germany} \\
\vspace{1.3cm}
{\bf Abstract}
\end{center}
\vspace{-0.1cm}
\normalsize
The differences are discussed between various next-to-leading order 
prescriptions for the QCD evolution of parton densities and structure 
functions. Their quantitative impact is understood to an accuracy of 
0.02\%. 
The uncertainties due to the freedom to choose
the renormalization and factorization scales are studied. The
quantitative consequences of the different uncertainties on the 
extraction of the strong coupling constant $\alpha_s$ from scaling 
violations in deep--inelastic scattering are estimated for the 
kinematic regime accessible at
 HERA.

\vspace{8mm}
\noindent
\normalsize
To appear in the Proceedings of the 
{\sf 1996 Zeuthen Workshop on Elementary Particle Theory: QCD and QED
in Higher Orders\/}, Rheinsberg, Germany, April~1996,
Nucl.\ Phys.\ {\bf B} (Proc.\ Suppl.).
 
\end{titlepage}

\begin{abstract}
\vspace*{5mm}
\noindent
The differences are discussed between various next-to-leading order 
prescriptions for the QCD evolution of parton densities and structure 
functions. Their quantitative impact is understood to an accuracy of 
0.02\%. 
The uncertainties due to the freedom to choose
the renormalization and factorization scales are studied. The
quantitative consequences of the different uncertainties on the 
extraction of the strong coupling constant $\alpha_s$ from scaling 
violations in deep--inelastic scattering are estimated for the 
kinematic regime accessible at
 HERA.
\vspace*{-1mm}
\end{abstract}
\maketitle
 
\section{Introduction}  
\label{sect1}

\noindent
Deeply inelastic scattering (DIS) has a long and successful history in 
the discovery and analysis of the substructure of the nucleon. It 
provides one of the cleanest tests of QCD. With the advent of HERA, 
the kinematic range
has been
extended to $Q^2$ values up to about $10^4\: {\rm GeV}^2$. Moreover in 
the Bjorken--variable $x$, the range down to values of $x\simeq 10^{-4}$
is now probed in the deep--inelastic regime. This extended kinematic
coverage allows for detailed measurements of the scaling violations of  
the structure function
$F_2(x,Q^2)$ and, consequently, of the strong coupling constant
$\alpha_s$. The statistical and systematic
experimental errors for such analyses have been estimated in 
refs.~\cite{BIMR,EXPA}.

The determination of $\alpha_s$ in analyses of DIS
data from high statistics experiments relies on perturbative QCD
expansions, implemented in complicated fitting procedures. Presently
the necessary theoretical ingredients are fully known only up to 
next-to-leading order (NLO). In this paper, we analyze the
theoretical ambiguities arising at this level of accuracy, and study 
their implications for the extraction of $\alpha_s(M_Z^2)$ from 
the scaling violation of the structure function $F_2$.

The paper is organized as follows. In Section~2 we give a
detailed discussion of the various approximations used in different
solutions of
the evolution equations. We
examine their quantitative
impact on
the predicted scaling violations of the
parton densities and structure
functions. 
Section 3 is devoted to the uncertainties originating from
the choice
of the renormalization and mass  factorization scales at this order
of the perturbative expansion. The implications of all these effects 
for the theoretical error $\Delta \alpha_s(M_Z^2)$ from QCD analyses 
of DIS scaling violations in the kinematic range at HERA are
presented in Section 4. 
Section 5 contains the conclusions.

\vspace{-1mm}

\section{Structure function evolution in NLO}  
\label{sect2}
 
\noindent
In this section the details of  various 
representations
 of
the  evolution equations 
in NLO and the renormalization group equation
determining $\alpha_s$
are
compared.  To keep the notation as compact and transparent as
possible, the evolution equations will be written in a generic manner, 
which covers the  non--singlet equations
as well as the coupled
singlet quark and gluon evolution equations. 
Explicit solutions  written down
in Sections 2.2 and 2.3 will apply literally only to the non--singlet 
cases. A numerical comparison of the different prescriptions 
is performed
in Section 2.4.

\vspace{3mm}
\subsection{Evolution equations for parton densities
and ${\boldmath\alpha_s(Q^2)}\!$}
\label{sect21}

\vspace{1mm}
\noindent
The evolution equations for the (twist-2) parton distributions 
$f(x,M^2)$ 
of the nucleon are given by
\bea
\label{evol1}
 \frac{\partial f(x,M^2)}{\partial \ln M^2 } &\!\! =\!\! &
 \left[ a_{s}(M^2)\, P_{0}(x) + a_s^2(M^2)\, P_1(x) \right. \nnn 
 & & \left. \mbox{} + O(a_s^3) \right] \otimes f(x,M^2)  \: .
\eea
As usual $x$ stands for the nucleon's momentum fraction carried by 
the partons, and $\otimes $ denotes the Mellin convolution,
\beq
\label{convx}
 A(x) \otimes B(x) = \int_x^1 \! \frac{dy}{y} \, A(y) \, B\! 
 \left(\frac{x}{y}\right) \: .
\eeq
$M$ represents the mass  factorization scale. In eq.~(\ref{evol1}) we
have already inserted the perturbative expansion of the splitting 
functions $P(x,M^2)$ in powers of the strong coupling constant 
$ \alpha_s(M^2) \equiv 4 \pi a_s(M^2)$. Only the first two expansion 
coefficients $P_0(x)$ and $P_1(x)$ are completely known so far.
Hence the solution of the evolution equations is presently possible up 
to NLO.

To this accuracy the scale dependence of $a_s(R^2)$ is governed by
\beq
\label{arun1}
\frac{\partial \, a_s(R^2)}{\partial \ln R^2 } =
- \beta_0 a_s^2(R^2) - \beta_1 a_s^3(R^2) + O(a_s^4) \: ,
\eeq
with the first two (scheme independent) coefficients of the QCD 
beta function given by $\beta_0 = 11 - 2 N_f/3 $ and $\beta_1 = 102 -
38 N_f/3 $. Here $N_f$ denotes the number of quark flavours. As already 
indicated in eq.~(\ref{evol1}), the renormalization scale $R$ is 
identified with the factorization scale in this section, $ R = M $. 
For the case that $ R $ and $M$ are chosen to be unequal see
Section~3.

Like
the solution of the parton evolution equation (\ref{evol1}),
discussed in detail below, also
the NLO running of $a_s$ can be treated in 
several ways, differing beyond the accuracy of the present 
approximation. First, eq.~(\ref{arun1}) can be simply solved by
numerical iteration starting from a reference scale $R_0 $, or
equivalently by using its integrated implicit form
\bea
\label{arun2}
\frac{1}{a_s(R^2)} &=& \frac{1}{a_s(R_0^2)}
- \beta_0 \ln \left( \frac {R^2}{R_0^2} \right) \\
 &+& \frac{\beta_1}{\beta_0}
 \ln \left \{ \frac{a_s(R^2) [\beta_0 + \beta_1 a_s(R_0^2)]}
                   {a_s(R^2_0) [\beta_0 + \beta_1 a_s(R^2)]}  \right \}.
\nonumber
\eea
Second, since in lowest order $ 1/a_s \sim \ln (R^2/R_0^2) $, one can
treat eq.~(\ref{arun1}) in the sense of a power expansion in $1/ \ln
(R^2/R_0^2) $, and keep only the powers to which the terms in 
eq.~(\ref{arun1})
beyond $\beta_1$ do not contribute.
Introducing the QCD
scale parameter $\Lambda $ instead of the reference value $a_s(R_0^2)$,
the result can be written as
\bea
\label{arun3}
 a_s(R^2) & \!\! =\!\! & \frac{1}{\beta_0 \ln(R^2/\Lambda^2)}
 - \frac{\beta_1}{\beta_0^3} \frac{\ln\, [\ln(R^2/\Lambda^2)]}
 {\ln^2 (R^2/\Lambda^2)}  \nnn
 & & \mbox{} + O \left( {\ln^{-3} (R^2/\Lambda^2)} \right) \: . 
\eea
Both these treatments, as well as slightly different analytic
approximations~\cite{DR}, e.g.\ by truncating in $1/a_s$ instead of 
$a_s$ as in eq.~(\ref{arun3}), have been widely used in the context of 
parton evolution. 

Let us now return to the evolution equation (\ref{evol1}).
This system of coupled integro--differential equations has been
treated in various manners. Aside from
  expansions in orthogonal
polynomials~\cite{ORTH}, which will not be discussed here, the 
principal options are to solve these equations in
$x$-space or to consider their
transformation using
Mellin-$N$ moments, 
\beq
\label{Mellin}
 A^{N} \equiv \int_0^1 \! dx \, x^{N-1} A(x) \: .
\eeq
In the latter case the convolution (\ref{convx}) reduces to the
product
\beq
\label{convn}
 [ A(x) \otimes B(x) ]^N =  A^N \, B^N \: ,
\eeq
and eq.~(\ref{evol1}) becomes a system of ordinary differential
equations at fixed $N$.

In many
analyses, eq.~(\ref{evol1}) has been solved numerically
in $x$-space, see e.g.~\cite{VO}--\cite{PZ},
without further reference to the power--series structure in $a_s$.
We will denote this approach, which is a direct solution of the
renormalization group equations in the case of mass factorization,
as prescription {\bf (A)} in the numerical comparisons in Section 2.4. 
On the other hand, the transformation of the evolution equations
to $N$-space~\cite{DFLM,GRV}
allows for further analytic developments to which we now turn.

\vspace{3mm}
\subsection{Analytic solutions and approximations$\!$} 
\label{sect22}

\vspace{1mm}
\noindent
The first step towards an analytic solution of the evolution equations
is to rewrite eq.~(\ref{evol1}) in terms of $a_s$:
\beq
\label{evol2}
\frac{\partial f^N(a_s)}{\partial a_s} = -\frac{a_s P_0^N + a_s^2 P_1^N 
 + O(a_s^3)}{\beta_0 a_s^2 + \beta_1 a_s^3 + O(a_s^4)} f^N(a_s) \, .
\eeq
To simplify the notation, the argument $M^2$ of $a_s$ is 
suppressed throughout this section. This formulation
is  still fully equivalent to eq.~(\ref{evol1}) in combination
with eq.~(\ref{arun2}).  The analytic
solution of eq.~(\ref{evol2}), which is possible in a closed form for 
the non--singlet cases,  or 
a numerical iterative treatment, can be
 used to cross-check the numerical accuracy of concrete $x$- and
$N$-space implementations,~see~ref.~\cite{BBNRVZ}.

Expanding the r.h.s.\ of eq.~(\ref{evol2}) into a power series in
$a_s$, one arrives at
\bea
\label{evol3}
 \lefteqn{ \frac{\partial f^N(a_s)}{\partial a_s} = } \\
  & & \hspace*{-6.5mm} 
  - \frac{1}{\beta_0 a_s} \big[ P_0^N + a_s (P_1^N - \frac{\beta_1}
 {\beta_0} P_0^N) + O(a_s^2) \big] f^N(a_s) \: . \nn
\eea
The derivation of eq.~(\ref{evol3})
consists of two steps: first one assumes that the
beta function is truncated for $1/a_s $ [and not for $a_s$ as in
eq.~(\ref{arun1})], which is of course equivalent to the order 
considered.  Then the $a_s^2$ term arising in the square bracket is 
discarded, since at the same order in $a_s$  also the presently 
unknown next-to-next-to-leading order (NNLO) splitting functions 
$P_2^N$ enter. Hence this term is beyond the order being considered
here. 
The direct solution of eq.~(\ref{evol3}) will be referred to as 
prescription {\bf (B)}~\cite{itsol}
 in the numerical comparisons below.

Keeping the power--series character of
the evolution equations,
the final step of the solution
then leads to
\bea
 \label{evol4}
 f^N(a_s) &\!\!\! =\!\! &\!\!\! \left[ 1 - \frac{a_s-a_0}{\beta_0} \Big(
 P_1^N - \frac{\beta_1}{\beta_0} P_0^N \Big) + O(a^2_s) \right] \nnn
 & & \cdot \left( \frac{a_0}{a_s} \right)^{P_0^N/\beta_0}
 f^N(a_0) 
\eea
in the non--singlet cases,
with $a_0 \equiv a_s(M_0^2)$. Here $M_0$ is the reference scale at 
which the non--perturbative initial distributions are specified. With
respect to the $a_s$ expansion, the notationally more involved
singlet--matrix solution behaves as eq.~(\ref{evol4}), see
ref.~\cite{FPE1}.
 As compared to
this final result, an iterative solution of eq.~(\ref{evol3})
generates also higher powers of the $P_1^N$, the
contribution of
which is
again  beyond NLO in the sense of an expansion in powers of
$a_s$.
In Section 2.4, we will refer to the  solution
(\ref{evol4}), which has 
practically been employed, for example, in refs.\ \cite{DFLM,GRV}, 
as prescription {\bf (C)}.

The Mellin--moment technique is a very powerful tool,
allowing the implementation of  various
kinds of approximations. As we will
mainly rely upon
this method in our subsequent numerical work, a short
reminder about
the transformation of the results back to $x$-space
is in order.
The  inverse Mellin transformation is performed
by a contour integral in the
complex $N$-plane, 
\beq
\label{minv}
 f(x,a_s) = \! \frac{1}{\pi} \int^{\infty}_0 \!\!\! dz \, {\rm Im} \,
 [ e^{i \phi} x^{-C} f^{N = C}(a_s) ] \, ,
\eeq
where $C = c + z e^{i \phi} $. 
Here all functions $f^N$
encountered
satisfy $(f^N)^* = f^{N^*}$.
The parameter
$c$ is chosen
 about one to two
units to the right of the rightmost singularity of $f^N$ -- at $N=1 \, 
(0)$ for singlet (non--singlet) cases. All singularities are
situated on the real axis for the NLO evolution, and $\phi \approx 
3\pi /4 $ \cite{GRV} can be safely used even down to extremely low $x$. 
The choice of $\phi > \pi /2$ leads to a faster convergence of the
integral (\ref{minv})
as $z \rightarrow \infty$. 
Using a chain of Gauss quadratures, a numerical accuracy
better than $10^{-5}$ is easily achieved.

\vspace{3mm}
\subsection{Structure function evolution near $Q_0^2$}
\label{sect23}

\vspace{1mm}
\noindent
The  structure function $F_2$ is obtained, for $ M^2
= R^2 = Q^2 $, from the (anti-) quark and gluon densities $f_q$ and 
$f_g$ considered in the previous sections by
\beq
\label{eqF2}
 F_2(x,Q^2) = \sum_{i=q,g} c_i[x,a_s(Q^2)] \otimes f_{i}(x,Q^2) \: .
\eeq
Here $c_{q,g}[x,a_s(Q^2)]$ denote the NLO
 coefficient functions
for quarks and gluons
\bea
\lefteqn{ c_i[x,a_s(Q^2)] = c_{0,i} \delta_{iq}\, \delta (1-x) +
a_s(Q^2) \, c_{1,i}(x) \, , \nn } \\
 & & 
\eea
where we identify the factorization scale $M$ by setting $M^2 = Q^2$.
If $Q^2$ is not too far from the reference scale $Q^2_0$, i.e.\ if the
evolution distance $L = \ln(Q^2/Q^2_0)$ is small, one may
Taylor expand eq.~(\ref{eqF2}) around $Q^2_0$.
In this way $F_2(x,Q^2)$ can be expressed in terms of $L$, the 
expansion coefficients 
 of the coefficient 
 and splitting functions,
$c_{i}(x)$ and $P_{kl}(x)$,
the parton densities
$f_{i}(x,Q^2_0) \equiv f_{i,0}(x)$, 
and the strong
coupling 
$a_s(Q_0^2) \equiv a_0$
at the initial scale.

The explicit result for this case will also be given for the
notationally simpler non--singlet structure functions only. 
We will again
switch to Mellin moments below. The approximation under consideration 
is most easily derived directly from eqs.~(\ref{evol1}) and 
(\ref{arun1}) without reference to their solutions discussed above.
First the $a_s$-equation (\ref{arun1}) is solved up to orders $a_0^3$ 
and $L^2$,
resulting in 
\beq
 a_s(Q^2) \, \simeq \, a_0 - a_0^2 \beta_0 L - a_0^3 \beta_1 L 
 + a_0^3 \beta_0^2 L^2 \: .
\eeq
The expansion of the r.h.s.\ of eq.~(\ref{eqF2}) in connection with 
eq.~(\ref{evol1}) is then, including terms  up to $O(a_0^2)$ and 
$O(L^2)$, given by 
\bea
\label{local}
 F^{N}(Q^2) & \!\!\simeq \!\!& \Big\{ 1 + a_0 [c_{1,q}^{N} + P_0^N L]
  \\ 
 & & \mbox{} + a_0^2\Big[ (P_1^N +P_0^N c_{1,q}^N -\beta_0 c_{1,q}^{N})
  L \nonumber \\
 & &\mbox{} \hspace{7mm} + \frac{1}{2} ([P_0^N]^2 -\beta_0 P_0^N) L^2
  \Big] \Big\} \, f^N_0 \, . \nonumber 
\eea
One may express this solution in terms of the  coefficient
function to order $a_s^2$, $C_q(a_0, Q^2/Q^2_0)$~\cite{COEFF}
as
\bea
 F^{N}(Q^2) & \!\simeq \! & \Bigl \{ (1 + a_0 \, c_{1,q}^{N}) + 
 \Bigl [C_q^N (a_0, Q^2/Q^2_0) \nonumber\\
 & & \mbox{} - C_q^N (a_0,1) \Bigr ] \Bigr \} \, f^{N}_0 \: .
\eea
Here the general form of the coefficient function
\begin{eqnarray}
C_q^N (a_s, Q^2/M^2) &=& c_{0,q} + a_s C_{1,q}^N(Q^2/M^2)
                               \nonumber\\
&+& a_s^2 C_{2,q}^N(Q^2/M^2) + O(a_s^3)
\nonumber\\
\end{eqnarray}
was used choosing $M^2 = Q^2_0$.
The local representation (\ref{local}) of the structure function 
evolution will be denoted as prescription {\bf (D)}~\cite{vnb}
  in the discussion
below.

\vspace{3mm}
\subsection{Numerical comparisons}
\label{sect24}

\vspace*{1mm}
\noindent
We now turn to the comparisons of the numerical results obtained by 
the prescriptions {\bf (A)--(D)} defined in the previous sections.
All these versions are  equivalent to NLO. Their comparison
gives  information about the uncertainty induced at the level of
possible approximations. The uncertainties due to different choices of 
the renormalization and factorization scales are discussed in Section~3.

The results will be shown for initial distributions which, although
representing a somewhat simplified input, incorporate all features
relevant to this study in a sufficiently realistic way. Specifically, 
we take in the \MSbar\ factorization scheme at $M_0^2 = 4 $ GeV$^2$:
%
\begin{eqnarray}
\label{input}
  xu_v \! &\! =\! & A_u x^{0.5} (1-x)^3, \:\: xd_v \, = \, A_d x^{0.5}
  (1-x)^4, \nonumber\\
  xS &\! =\! & \Sigma \, - \, xu_v \, - \, xd_v~~~~ = \, A_S x^{-0.2}
  (1-x)^7, \nonumber\\
  xg &\! =\! & A_g x^{-0.2} (1-x)^5, \:\: xc \, = \, x\bar{c} \: = \: 0
  \: .  \nonumber\\
\end{eqnarray}
The evolution is performed for four massless flavours,
using $\Lambda_{\overline{\rm MS}}(N_f\! =\! 4) = 250$ MeV in eq.\
(\ref{arun3}). The SU(3)--symmetric  sea is assumed to carry 15\% of 
the nucleon's momentum at the input scale, and the remaining 
coefficients are fixed by sum rules.
Also the results on $F_2$ will be shown employing $M^2 = R^2 = Q^2$.
The massless charm evolution does not yield a correct representation
of $F_2^{c\overline{c}}(x,Q^2)$.
However, the conclusions of this
investigation are not substantially affected by this simplification.

In Figure 1 the singlet quark and gluon densities  obtained from the
solutions {\bf (A)--(C)} are compared, after an
evolution to $M^2 = 100$ 
GeV$^2$ and $10^4$ GeV$^2$. To display the differences
clearly, the curves have been normalized to the values obtained for
case {\bf (C)}. Between the input scale and the
highest scale considered in the figure, $x\Sigma $ and $xg$ increase 
(decrease) by more than an order of magnitude at very low (high) values
of $x$, respectively.

\begin{figure}[thb]
\vspace*{-9mm}
\begin{center}
\mbox{\hspace*{-4mm}\epsfig{file=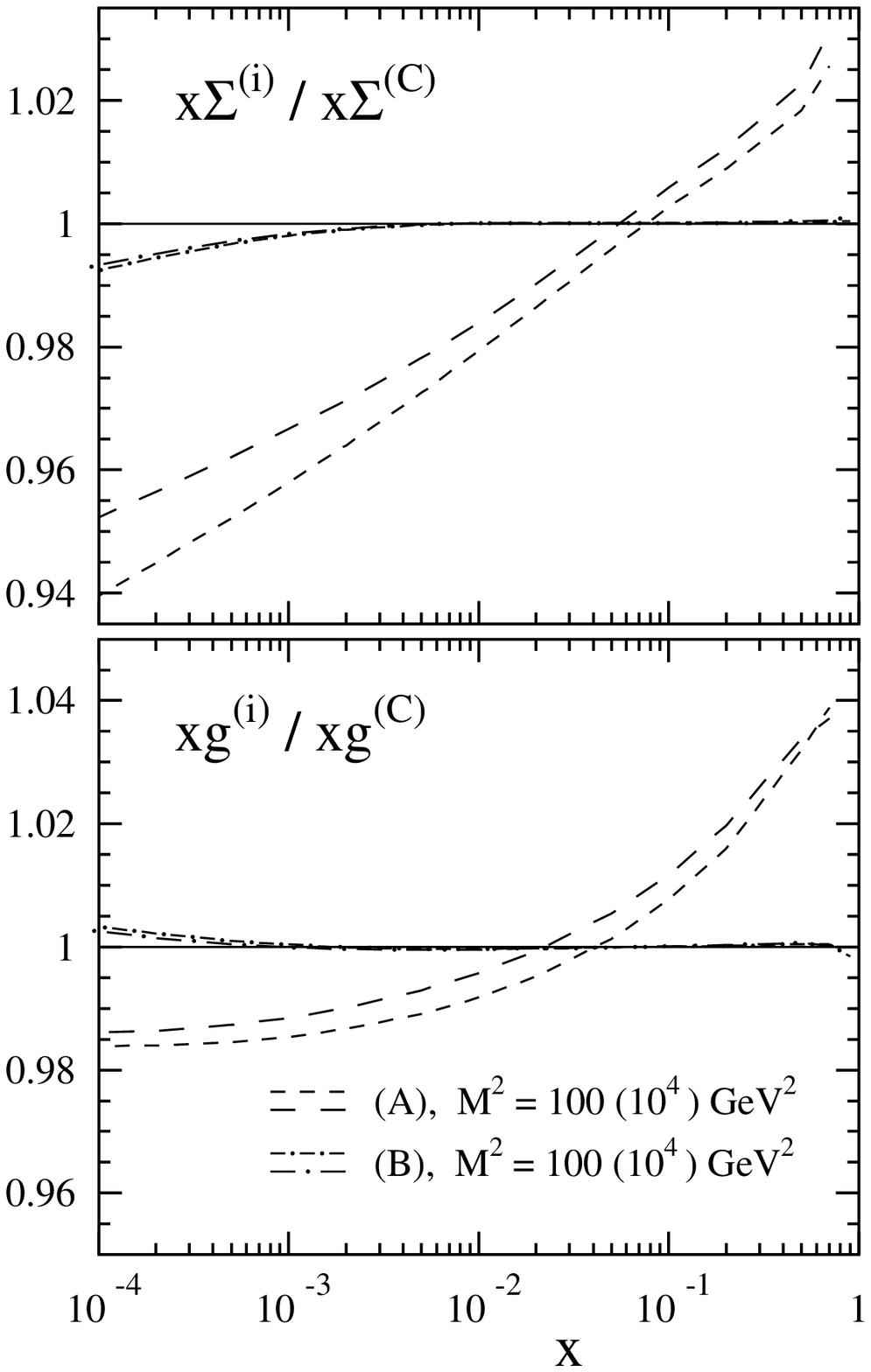,width=8.4cm}}
\vspace*{-10mm}
\end{center}
{\sf {\bf Figure 1:}~~A comparison of the singlet quark and gluon
 densities, $x\Sigma $ and $xg$, as evolved using the prescriptions 
 {\bf (A, B)} discussed in the text, normalized to {\bf (C)}
 for the initial conditions
 (\ref{input}). The results are displayed at two representative
 choices of the factorization and renormalization scale $M$.}
 \vspace{-2mm}
\end{figure}
The difference between the results based on eq.~(\ref
{evol3}), {\bf (B)} and {\bf (C)}, is very small for $ 10^{-3}
< x < 0.9 $, and reaches at most
the order of
1\% for smaller $x$. This small difference
is due to the iterated $P_1$ terms in the solution of eq.~(\ref{evol3}).
On the other hand, the deviations of the 
evolution {\bf (A)} based upon eq.~(\ref{evol1}) from these results are
rather large.  These offsets are not related to any numerical
inaccuracies, since those are under control to the level of  
0.02\%~\cite{BBNRVZ}.
Instead these differences are due to both of the two expansion steps 
discussed below eq.~(\ref{evol3}), with the first one, the re--expansion
of the beta  function, being somewhat more important numerically.
As will be shown in Section~4,
these differences in $F_2$ result in   a shift
of $\alpha_s(M_Z^2)$ by 0.003 under the conditions at HERA.

\begin{figure}[thb]
\vspace{-9mm}
\begin{center}
\mbox{\hspace*{-3mm}\epsfig{file=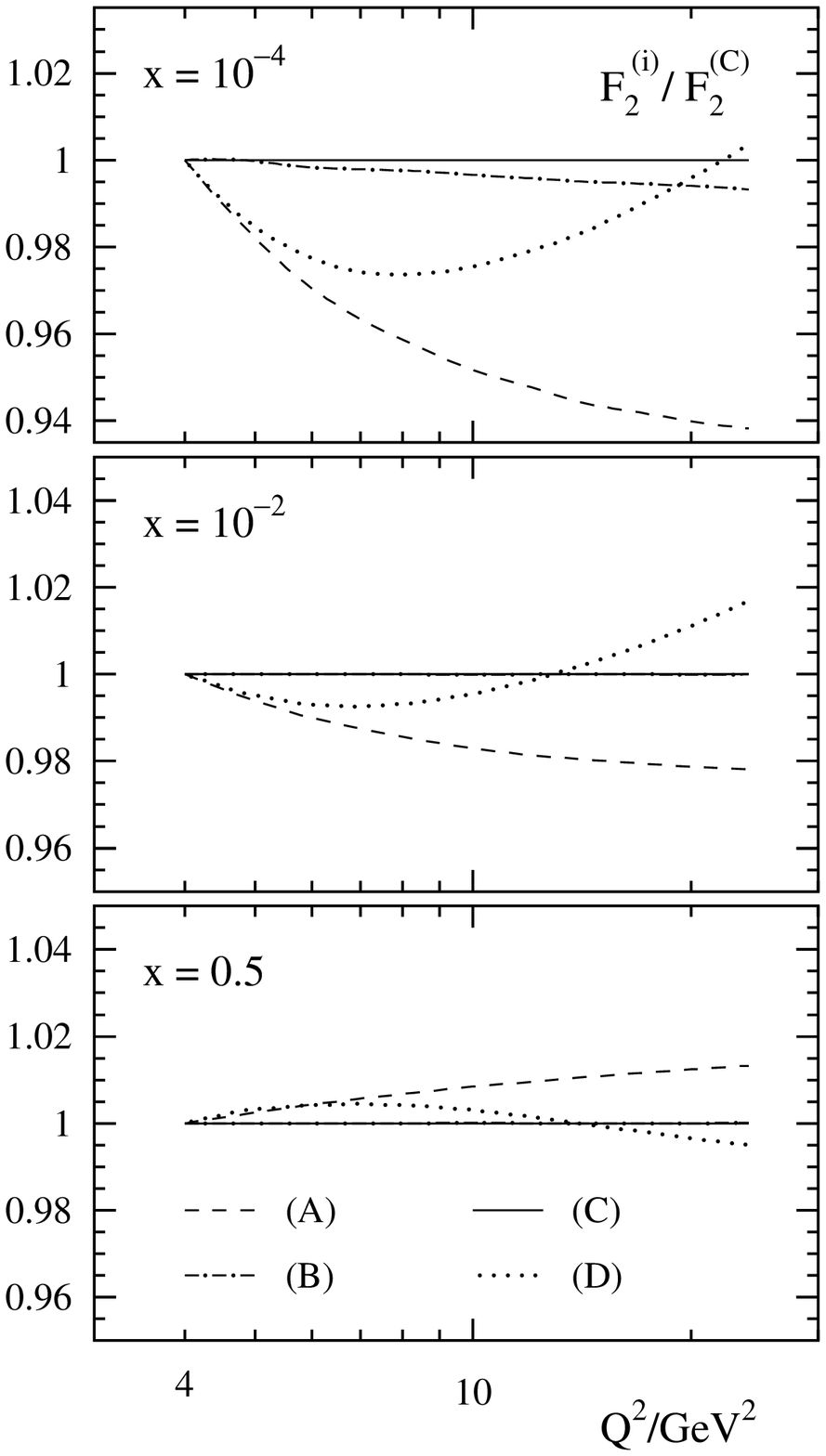,width=8.2cm}}
\vspace*{-10mm}
\end{center}
{\sf {\bf Figure 2:}~~A comparison of the $Q^2$ evolutions of the proton
 structure function $F_2(x,Q^2)$ as obtained using the prescriptions
 {\bf (A, B, D)}, normalized to {\bf (C)}
 close to the reference scale $Q_0^2 = 4 \: {\rm
 GeV}^2$.} 
 \vspace{-7mm}
\end{figure}

Figure 2 shows how the offset between the solutions {\bf (A)} and 
{\bf (B,C)} develops, for three typical values of $x$, with increasing 
$Q^2$ in the structure function $F_2$. The result of the local
representation {\bf (D)}, representing an
expansion valid
for $L = \ln(Q^2/Q^2_0)  \ll 1$, is also given here.
For small values of $L$
it agrees with  scheme {\bf (A)}.
The
figure indicates that the difference, observed in the previous 
comparison,
builds up very quickly in the vicinity of $Q_0^2$, and then stays at
about the
 same level above $Q^2 \approx 20 \mbox{ GeV}^2$. This
behaviour is readily explained by the large  size of $\alpha_s$
at scales around $4 \mbox{ GeV}^2$: $\alpha_s^2 $ changes by
almost a factor of two between 4 and 20 GeV$^2$.

\section{Renormalization and Factorization Scale$\! $ Uncertainties}
\label{sect3}

\noindent
The renormalization scale dependence of non-singlet parton densities
$f^{N}$ is determined by
\bea
\label{eqn:rendep}
\lefteqn{ {\partial f^{N}(M^2,R^2) \over \partial \ln M^2 } = 
 a_s(R^2) \left[ P_0^N + a_s(R^2) P_1^N
\right. }  \\
 & & \left.  + a_s(R^2)
\beta_0  P_0^N \ln \left( {R^2 \over M^2}
  \right)  \right] f^{N}(M^2,R^2) \: . \nonumber
\eea
Here the values of $a_s$ at the factorization scale $M^2$ and 
renormalization scale $R^2$ are, to NLO, related by
\beq
 a_s(M^2) = a_s(R^2) \left [1 + \beta_0 a_s(R^2) \ln \left(
 \frac{R^2}{M^2}\right)\right] .
 \label{eq:rela}
\eeq
In fact, one can resum the terms in front of the LO splitting functions,
$P_0^N$, in eq.~(\ref{eqn:rendep}) using the relation (\ref{eq:rela}). 
Thus to NLO accuracy
the parton densities can be expressed as a function of a single scale.

The structure functions $F^N(Q^2)$ are represented by the solution of 
eq.~(\ref{eqn:rendep}) convoluted with the coefficient functions 
according to
\beq
 F^{N}(Q^2) = \bigl[ 1 + a_s(R^2) c_{1,q}^N \bigr] f^{N}(Q^2,R^2) \: .
 \label{FNS1}
\eeq
For the study of the renormalization scale dependence we identify the
mass factorization scale $M^2 = Q^2$.

The factorization scale dependence of the structure function $F^N$
is described by
\bea
\label{FNS2}
 \lefteqn{ F^{N}(Q^2) = } \\ & & \hspace*{-3mm} 
 \Bigl [ 1 + a_s(M^2) \Bigl ( c_{1,q}^N + P_0^N \ln \left( 
 \frac{Q^2}{M^2} \right) \Bigr) \Bigr] f^{N}(M^2) \, , \nonumber
\eea
where $f^{N}$ is the solution of
\bea
 \lefteqn{ 
 {\partial f^N(M^2) \over \partial \ln M^2 } = a_s(M^2) \Bigl[ 
  P_0^N + a_s(M^2) P_1^N \Bigr]  f^{N}(M^2) . } \nonumber \\
  & &
\label{eqn:facdep}
\eea
Here the renormalization scale is fixed by $R^2 = M^2$.

Eqs.~(\ref{FNS1}) and (\ref{FNS2}) are independent
of the choices of the scales $R$ and $M$, respectively, up to NLO.
A further improvement of the scale stability, up to $O(\alpha_s^3)$, 
can be obtained extending the above analysis to 3--loop order, if 
the corresponding splitting functions become available.  To
derive an estimate of the quantitative importance of the present 
scale dependences, $R^2$ and $M^2$ will be varied in the range 
between $Q^2/4$ and $4 Q^2$ in Section~4.

\vspace*{6.4cm}
\includegraphics{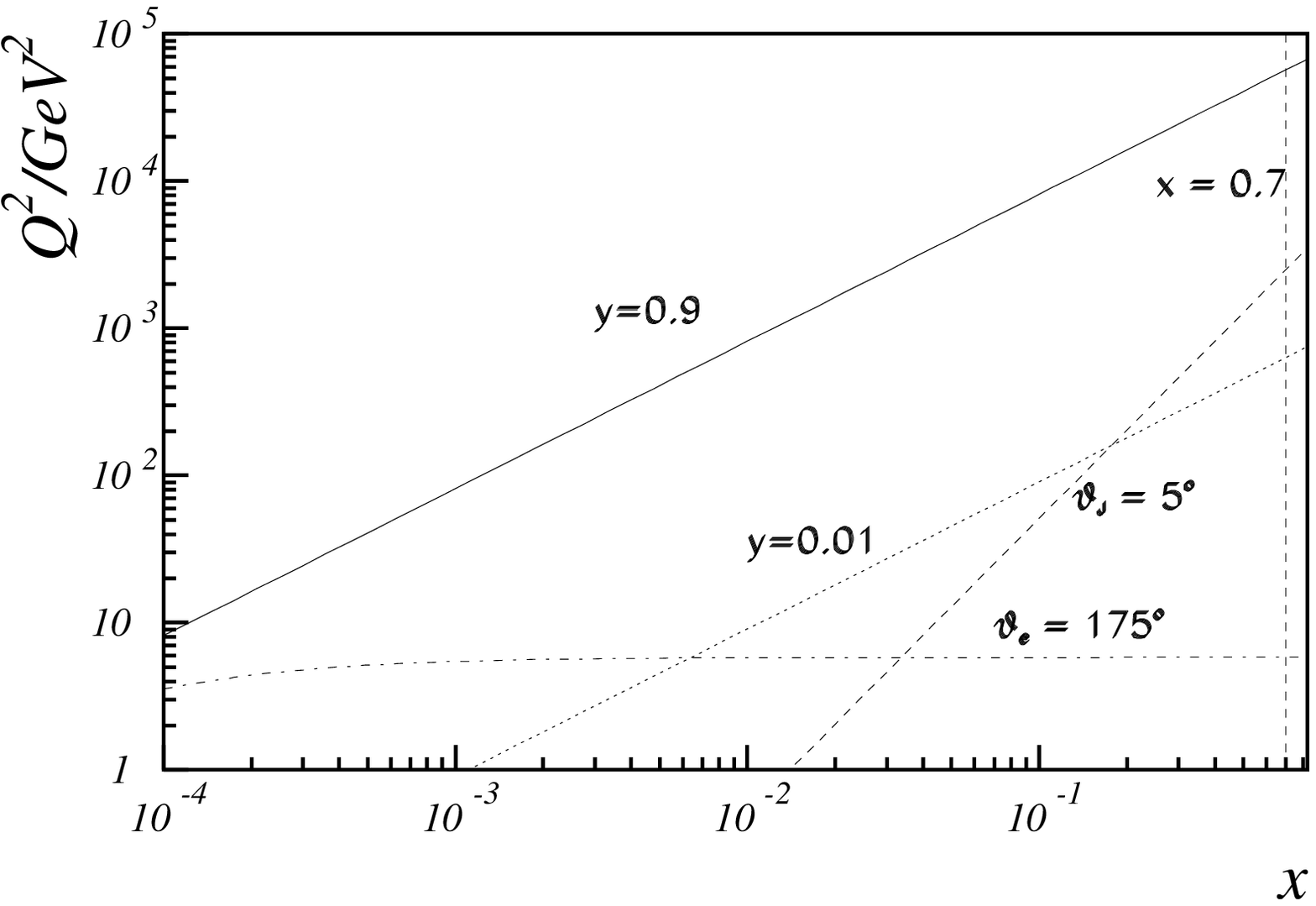}
\vspace{0.2cm}
\noindent
{\sf {\bf Figure~3:}~~The kinematic range for DIS measurements
 at HERA used for the determination of the theoretical uncertainties
 of $\alpha_s(M_Z^2)$ in Section 4.}
\vspace{0.4cm}

\section{Estimation of the Theoretical Error $\Delta \alpha_s(M_Z^2)$}
\label{sect4}

\noindent
The necessity of using
perturbative solutions of the renormalization group equations in
representing the observables implies 
theoretical  errors in the determination  of $\alpha_s$ in all
experimental analyses. 
It is convenient to compare
them at a common reference
scale for which we choose $\mu^2 = M_Z^2$. Furthermore we
 refer
to the $\overline{\rm MS}$ renormalization
scheme below.
Recent $\alpha_s$ measurements were compiled
in~\cite{bethke,jb}.  
The most precise
results obtained  show that the values
of $\alpha_s(M_Z^2)$ as determined
in deep inelastic scattering
and in
$e^+e^-$  annihilation  seem to differ.
The central value determined by the
DIS data is~\cite{jb}
\begin{equation}
\alpha_s(M_Z^2) = 0.112 \pm 0.004,
\end{equation}
whereas in the $e^+e^-$ experiments
\begin{equation}
\alpha_s(M_Z^2) = 0.121 \pm 0.004
\end{equation}
is obtained.
Future high statistics measurements of $\alpha_s$ from deep inelastic
scattering at HERA may help to
resolve this difference. For the interpretation of the data, a careful
treatment of the theoretical errors is required.
In the following we study possible sources of theoretical uncertainty
in the framework of the
NLO evolution. They include:
\begin{itemize}
\item[1)]
 The effect arising from the different representations of $\alpha_s$ 
 given in eqs.\ (\ref{arun2}) and (\ref{arun3}).
\item[2)]
 The offsets originating in the different NLO prescriptions 
 {\bf (A)--(C)}, respectively based upon eqs.\ (\ref{evol1}),
 (\ref{evol3}) and (\ref{evol4}), for the structure function 
 evolution.
\item[3)]
 The theoretical uncertainties due to the freedom of choice
 for the renormalization and mass factorization scales.
\end{itemize}
To gauge the effect of each source of uncertainty, a reference data
set was constructed for which the different displacements were measured
using the $\chi^2$ method in the kinematic   regime of HERA.
The different cuts used are illustrated in figure~3.
The value of $\alpha_s(M_Z^2)$ was taken to be
$0.112$
fixing $N_f = 4$ in
the entire range of $Q^2$, taking prescription {\bf (A)} for the
evolution, and using eq.~(\ref{arun3}) for the description of $a_s(Q^2)$.

As outlined in Section~2 the solution of the evolution equations
either using eq.~(\ref{evol1}) or 
eq.~(\ref{evol3}) lead to different results.
In the determination of $\alpha_s(M_Z^2)$ a relative
shift about 0.003 is implied comparing both approaches. A similar
difference has been observed between various fits in the analysis of
the BCDMS $F_2^{\rm NS}$ data~\cite{BCDMS}, see Table 1.

The use of either the representation for $a_s$ as given in 
eq.~(\ref{arun3}) or the complete solution induces a
shift of $\Delta \alpha_s(M_Z^2)$ of 0.001 or less. A much smaller
difference  of 0.0001 in $\alpha_s(M_Z^2)$
is obtained comparing fits using a fully iterative solution
(\ref{evol3}) of the
evolution equation with one based on the expansion (\ref{evol4}).

The largest contributions to the theoretical error of $\alpha_s(M_Z^2)$
are due to the renormalization and factorization scale uncertainties.
In the present analysis,
we vary each scale separately in the range $Q^2/4$ to $4 Q^2$.
The effects are particularly strong 
if no further cuts on $Q^2$ are applied, see Table~2. If only the $Q^2$
range above 50~GeV$^2$ is considered, the factorization scale
variation in the above range results in   $\Delta \alpha_s(M_Z^2)|_M
= \pm 0.003$, while the corresponding error for the renormalization
scale variation is estimated to be 
\\[0.1cm]
$\Delta \alpha_s(M_Z^2)|_R = {\footnotesize
\begin{array}{c}+0.004\\-0.006\end{array}}$.

\section{Conclusions}
\label{sect5}

\noindent
Various prescriptions for the NLO evolution of DIS structure functions
have been discussed, and their quantitative differences were studied.
The resulting {\it theoretical} uncertainties are rather large. At
$x \simeq 10^{-4}$, e.g., the spread reaches about 6\% for the
proton structure function $F_2(x,Q^2)$ for identical inputs. A shift
of the central value of $\alpha_s(M_Z^2)$ by about 0.003 is observed, 
when the same data in the HERA kinematic   region are fitted with
programs using the most differing prescriptions. 
This is to be 
contrasted to the offsets induced by the various {\it numerical} 
procedures. They are  under control between five independent
implementations~\cite{BOT,PZ,GRV,itsol,vnb}
of $F_2(x,Q^2)$ at the level of
0.02\%, see ref.~\cite{BBNRVZ}.

The dependence 
of the evolution of $F_2(x,Q^2)$
on the choice of the renormalization and mass  factorization scales
$R$ and $M$
has been investigated.
 By varying both scales independently within the range
$Q^2/4$ to $4 Q^2$, the theoretical error $\Delta \alpha_s(M_Z^2)$ of 
future high statistics $\alpha_s$ measurements from $F_2$ scaling
violations at HERA has been estimated. If a  $Q^2$ cut of 50 GeV$^2$
is applied, at NLO it amounts to  
$$\Delta \alpha_s(M_Z^2) = \small \left.
  \begin{array}{c}+0.004\\-0.006\end{array}\right|_{R}  \left.
  \begin{array}{c}-0.003\\+0.003\end{array}\right|_{M} .
$$
A significant reduction of the theoretical uncertainties can be
expected  if the presently unknown 3--loop splitting functions become
available.  

\vspace{5mm}
\noindent
{\bf Acknowledgements~:}~This work was supported on part by the EC Network
`Human Capital and Mobility' under contract No.\ CHRX--CT923--0004 and by
the German Federal Ministry for Research and Technology (BMBF) under
No.\ 05 7WZ91P (0).

\newpage
\onecolumn
\begin{center}
\vspace{-10mm}
\begin{tabular}{||l||c|c|c||}
\hline \hline
 & & & \\[-0.3cm]
\multicolumn{1}{||l||}{Program} &
\multicolumn{1}{c|} {$\Lambda_{{\overline
{\rm MS}}}$/MeV, $N_f = 4$} &
\multicolumn{1}{c|}{$\alpha_s(M_Z^2)$} &
\multicolumn{1}{c||}{$\chi^2$/d.o.f.} \\
 & & & \\[-0.3cm] \hline \hline
 & & & \\[-0.3cm]
 Krivokhizhin {\it et al.} \cite{KR}     &  $231 \pm 20$ & $0.1115 \pm
 0.0015$ & $176/152$  \\
 Virchaux, Ouraou \cite{VO}              &  $235 \pm 20$ & $0.1118 \pm
 0.0015$ & $178/150$  \\
 Gonz\'alez-Arroyo {\it et al.} \cite{GA}  &  $220 \pm 20$ & $0.1107 \pm
 0.0016$ & $180/151$  \\
 Abbott {\it et al.} \cite{AB}           &  $233 \pm 20$ & $0.1117 \pm
 0.0016$ & $198/151$ \\
 Furmanski, Petronzio \cite{FPE}         &  $270 \pm 25$ & $0.1144 \pm
 0.0016$ & $181/152$ \\
\hline \hline
\end{tabular}
\end{center}

\vspace*{1cm}
\noindent
{\sf {\bf Table~1:}~~Comparison of the QCD scale parameter
  $\Lambda_{ \overline {\rm MS}}$ obtained
 from NLO fits to the BCDMS $F_2^{\rm {NS}}$ data employing different
 evolution programs, taken from ref.~\cite{BCDMS}. The resulting values for 
 $\alpha_s(M_Z^2)$ have been computed, using eq.~(\ref{arun3}) and a 
 bottom threshold of $m_b   = 4.5~{\rm GeV} $.}


\begin{center}

\vspace{10mm}
\begin{tabular}{||c||c|c|c|c||}
\hline \hline
 & & & & \\[-0.3cm]
\multicolumn{1}{||l||}{Cut (GeV$^2$)} &
\multicolumn{1}{c|} {$M^2 = Q^2/4$} &
\multicolumn{1}{c|} {$M^2 = Q^2/2$} &
\multicolumn{1}{c|} {$M^2 = 2 Q^2$} &
\multicolumn{1}{c||} {$M^2 = 4 Q^2$} \\
 & & & & \\[-0.3cm] \hline \hline
 & & & & \\[-0.3cm]
 4  &        -- &       --  & $-0.0068$ & $-0.012 $ \\
 20 & $+0.0067$ & $+0.0029$ & $-0.0024$ & $-0.0044$ \\
 50 & $+0.0032$ & $+0.0015$ & $-0.0015$ & $-0.0029$ \\
\hline \hline
 & & & & \\[-0.3cm]
\multicolumn{1}{||l||}{Cut (GeV$^2$)} &
\multicolumn{1}{c|} {$R^2 = Q^2/4$} &
\multicolumn{1}{c|} {$R^2 = Q^2/2$} &
\multicolumn{1}{c|} {$R^2 = 2 Q^2$} &
\multicolumn{1}{c||} {$R^2 = 4 Q^2$} \\
 & & & & \\[-0.3cm] \hline \hline
 & & & & \\[-0.3cm]
 4  & $-0.0076$ & $-0.0037$ & $+0.0032$ & $+0.0059$ \\
 20 & $-0.0067$ & $-0.0032$ & $+0.0027$ & $+0.0049$ \\
 50 & $-0.0061$ & $-0.0028$ & $+0.0023$ & $+0.0042$ \\
\hline \hline
 & & & & \\[-0.3cm]
\multicolumn{1}{||l||}{Cut (GeV$^2$)} &
\multicolumn{1}{c|} {{\bf (B)},~eq.~(\ref{evol3})} &
\multicolumn{1}{c|} {{\bf (C)},~eq.~(\ref{evol4}) } &
\multicolumn{1}{c|} {$\alpha_s$,~eq.~(\ref{arun2})} &
\multicolumn{1}{c||} {} \\
 & & & & \\[-0.3cm] \hline \hline
 & & & & \\[-0.3cm]
 4  & $-0.0027$ & $-0.0001$ & $+0.0010$ &  \\
 20 & $-0.0025$ & $-0.0001$ & $+0.0009$ &  \\
 50 & $-0.0023$ & $-0.0001$ & $+0.0008$ &  \\
\hline \hline

\end{tabular}
\end{center}

\vspace*{1cm}
\noindent
{\sf {\bf Table~2:}~~The theoretical shifts on $\alpha_s
 (M_Z^2)$, from scale variations as well as from different NLO
 evolution and $\alpha_s$ prescriptions, for $F_2$ data in the HERA
 kinematic   range (see Figure~3).
 The reference data set was generated using
prescription {\bf (A)} for the evolution
 equations, eq.~(\ref {arun3}) for $\alpha_s$, and imposing
 $M^2 = R^2 = Q^2$.}

\vspace{-1mm}
\newpage
\twocolumn

\end{document}